\documentclass[journal,10pt]{IEEEtran} 
\usepackage[figurename=Fig.]{caption}
\usepackage{tikz}
\usetikzlibrary{calc, positioning, shapes.geometric, decorations.pathreplacing}
\usepackage{array}
\usepackage{verbatim}
\usetikzlibrary{calc, shapes, arrows, positioning}
\usepackage{authblk}
\usepackage{blindtext}
\usepackage{ifpdf}
\usepackage{graphicx}
\usepackage{tabulary}
\usepackage{amsmath,amsthm,amssymb}
\usepackage{amsmath}
\usepackage{kantlipsum}
\allowdisplaybreaks
\usepackage{cleveref}
\usepackage{lipsum}%
\usepackage{optidef}
\usepackage[english]{babel} 
\theoremstyle{plain}
\newtheorem{theorem}{Theorem}

\crefname{theorem}{Theorem}{theorem}
\crefname{lemma}{Lemma}{Lemmas}

\usepackage{cite}
\usepackage{dsfont}
\usepackage[justification=centering]{caption}
\usepackage{color}
\usepackage{algorithm}
\usepackage{algorithmicx, algpseudocode}
\usepackage{etoolbox}
\usepackage{subcaption}
\usepackage{balance}
\usepackage{empheq,etoolbox}
\usepackage[utf8]{inputenc}
\usepackage{multirow}
\usepackage{float}

\usepackage{amssymb}
\usepackage{pifont}
\usepackage[hmargin=1.27cm,top=1.3cm,bottom=1.3cm]{geometry}

\usepackage{physics}

\usepackage{babel}
\floatstyle{plaintop}
\restylefloat{table}
\patchcmd{\subequations}
\usepackage{lipsum} 
\allowdisplaybreaks

\tikzset{brace/.style={decorate, decoration={brace}},
 brace mirrored/.style={decorate, decoration={brace,mirror}},
}

\newcounter{brace}
\newcounter{arrow}
\begin{document}
 \captionsetup[figure]{name={Fig.},labelsep=period}

\title{Robust Resource Allocation for Pinching-Antenna Systems under Imperfect CSI}

\bstctlcite{IEEEexample:BSTcontrol}
\author{Ming Zeng, Xianbin Wang, \textit{Fellow, IEEE}, Yuanwei Liu, \textit{Fellow, IEEE}, Zhiguo Ding, \textit{Fellow, IEEE}, George K. Karagiannidis, \textit{Fellow, IEEE}, and H. Vincent Poor, \textit{Life Fellow, IEEE}
    \thanks{M. Zeng is with Laval University, Quebec City, Canada (email: ming.zeng@gel.ulaval.ca).}

    



    \thanks{X. Wang is with Western University, London, Canada (e-mail: xianbin.wang@uwo.ca).}


    \thanks{Y. Liu is with The University of Hong Kong, Hong Kong (email: yuanwei@hku.hk).}




\thanks{Z. Ding is with The University of Manchester, Manchester, UK. (e-mail:zhiguo.ding@ieee.org).}

\thanks{G. K. Karagiannidis is with Aristotle University of Thessaloniki, Thessaloniki, Greece (e-mail: geokarag@auth.gr).}

\thanks{H. V. Poor is with Princeton University, Princeton, NJ 08544 USA (e-mail: poor@princeton.edu).}
    }
\maketitle

\begin{abstract}
Pinching-antenna technology has lately showcased its promising capability for reconfiguring wireless propagation environments, especially in high-frequency communication systems like millimeter-wave and terahertz bands. By dynamically placing the antenna over a dielectric waveguide, line-of-sight (LoS) connections can be made to significantly improve system performance.  Although recent research have illustrated the advantages of pinching-antenna-assisted designs, they mainly presuppose complete knowledge of user locations—an impractical assumption in real-world systems. 
To address this issue, the robust resource allocation in a multi-user pinching antenna downlink system with uncertain user positions is investigated, aiming to minimize total transmit power while satisfying individual outage probability constraints. First, we address the single-user case, deriving the optimal pinching antenna position and obtaining the corresponding power allocation using a bisection method combined with geometric analysis. We then extend this solution to the multi-user case. In this case, we optimize the pinching antenna position using a particle swarm optimization (PSO) algorithm to handle the resulting non-convex and non-differentiable optimization problem. Simulation results demonstrate that the proposed scheme outperforms conventional fixed-antenna systems and validate the effectiveness of the PSO-based antenna placement strategy under location uncertainty.

\end{abstract}

\IEEEpeerreviewmaketitle

\section{Introduction} 
Recently, pinching antennas have emerged as a promising technology for reconfiguring wireless propagation environments \cite{Atsushi_22, yang2025, liu2025pinching}. In these systems, the antenna can be positioned along a dielectric waveguide to establish line-of-sight (LoS) communication links. This dynamic reconfigurability significantly improves system performance, especially in high-frequency bands, such as millimeter wave and terahertz frequencies. Studies have explored various deployment scenarios and consistently demonstrated the superiority of pinching-antenna-based systems over conventional fixed-antenna architectures.

Resource allocation (RA) design for pinching-antenna-assisted networks is gaining attention due to the crucial role that RA plays in achieving the full potential of such systems \cite{zeng2025_WCM}.  In pinching-antenna topologies, antenna location optimization is a crucial design factor, in contrast to fixed-antenna systems, where antenna positions are fixed.  However, complex joint optimization frameworks are required since this optimization is frequently closely associated with other wireless resources including time, frequency, and power.  The system goal determines the best RA approach.  Sum-rate maximization, for example, has been investigated in both uplink (\cite{Li25, Zeng_COMML25}) and downlink (\cite{ding2024, xie2025, wang2024, hu2025}) contexts.  Energy-efficient RA techniques have also been proposed for both uplink situations \cite{zeng2025EE} and downlink scenarios \cite{zeng2025EETVT, xie2025graph} in order to balance throughput and energy consumption.  Additionally, max-min rate optimization was discussed in \cite{Tegos_2025}, and total power minimization was examined in \cite{fu2025, wang2025modeling}.

 The majority of the previously listed works, however, make the strong assumption that the base station knows the user's coordinates perfectly. This assumption may not hold true in practice because positioning errors are unavoidable.  We examine a more practical multi-user downlink situation with the aid of a pinching antenna in order to overcome this constraint. In this scenario, the actual user location is situated within a circular uncertainty region that is centered at the estimated position.  Our goal under this paradigm is to minimize the overall transmit power while guaranteeing that each customer has a maximum outage probability constraint.

 We first examine the single-user scenario in order to address this non-convex problem with probabilistic constraints.  In this case, a bisection method combined with geometric analysis is used to solve the related power minimization problem and determine the ideal antenna site.  Under a fixed antenna position, the resulting power distribution technique can be expanded to the general multi-user scenario.  The particle swarm optimization (PSO) algorithm, which works well for non-differentiable and non-convex problems, is then used to optimize the pinching antenna position.
 The results of the simulation show that the suggested pinching-antenna-based scheme uses a lot less power than traditional fixed-antenna architectures.  Additionally, under location uncertainty, the PSO algorithm works well for figuring out the best antenna placement.

\begin{figure}[ht!]
\centering
\includegraphics[width=1\linewidth]{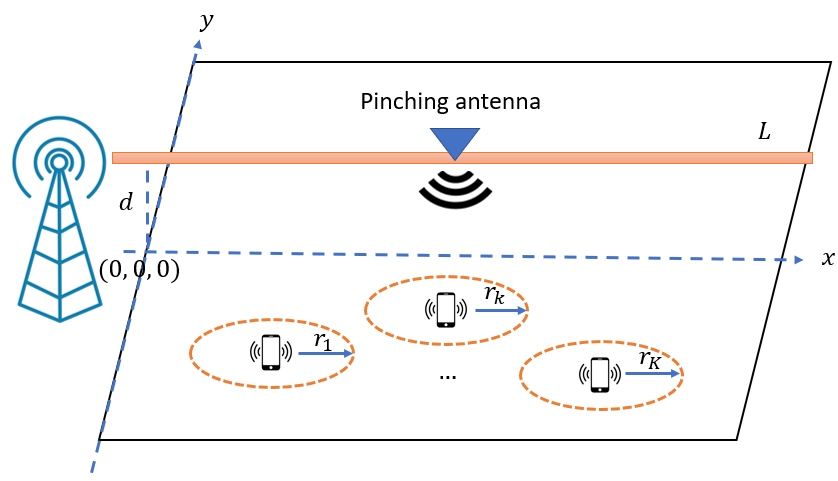}
\caption{A downlink pinching-antenna system with location uncertainty, $r_k$ is the radius of the uncertainty region.} 
\label{fig:Low_2}
\end{figure}

\section{System Model and Problem Formulation}
\subsection{System Model}
As shown in Fig. 1, we consider a downlink communication scenario in which an access point (AP) transmits signals to multiple single-antenna users via a pinching antenna deployed along a dielectric waveguide. To clearly represent the system geometry, a three-dimensional Cartesian coordinate system is adopted. The AP is positioned at the origin, i.e., at coordinates $ (0,0,0) $ meters. The dielectric waveguide is aligned along the $x$-axis, elevated at a height of $d$ meters above the ground plane, and has a total length of $L$ meters.

The users are randomly distributed within a rectangular region on the $xy$-plane. Let $K$ denote the total number of users. The estimated location of user $k$ is represented by $\Phi_k = (x_k, y_k, 0)$, for all $k \in \{1, \dots, K\}$. Unlike conventional works that assume perfect knowledge of $(x_k, y_k)$ at the AP, we consider a more practical scenario where user location information is imperfect. Specifically, while the AP is aware of the estimated coordinates $(x_k, y_k)$, the actual location of user $k$ may deviate within a circular uncertainty region centered at $(x_k, y_k)$ with radius $r_k$. The actual position of user $k$, denoted by $\hat{\Phi}_k = (\hat{x}_k, \hat{y}_k, 0)$ meters, satisfies the condition $(\hat{x}_k - x_k)^2 + (\hat{y}_k - y_k)^2 \leq r_k^2$ and $\hat{x}_k \geq 0$, ensuring the user remains on the same side of the waveguide. The user location is assumed to follow a uniform distribution within this circular region. Clearly, the smaller the value of $r_k$, the lower the uncertainty in the user’s position. The case $r_k = 0$ corresponds to perfect location knowledge.

In line with prior studies such as \cite{ding2024, xie2025}, we employ time-division multiple access (TDMA) with equal time allocation among users. Accordingly, user $k$ is served during the $k$-th time slot. To reduce system complexity and avoid frequent adjustment of antenna position, we assume that all users share a common pinching antenna position, denoted by $\Phi^{\text{Pin}} = (x^{\text{Pin}}, 0, d)$, where $x^{\text{Pin}} \in [0, L]$ \cite{Tegos_2025}. Considering only the dominant LoS channel component, the achievable data rate for user $k$ during its allocated time slot is computed using the free-space propagation model, as in \cite{ding2024, xie2025}

\begin{equation}
    R_k=\log_2 \left(1+ \frac{\eta P_k}{  \abs{\hat{\Phi}_k - \Phi^{\text{Pin}}}^2 \sigma_k^2} \right),
\end{equation}
where $\eta= \frac{\lambda^2}{16 \pi^2 }$ with $\lambda$ denoting the wavelength of the carrier frequency in free space \cite{Tegos_2025}. Additionally, $P_k$ represents the transmission power allocated to user $k$, while $\sigma_k^2$ denotes the noise power at user $k$.

Due to the uncertainty in user location, the actual user position $\hat{\Phi}_k$ is modeled as a random variable, and consequently, the achievable rate $R_k$ also becomes a random variable. A transmission outage is said to occur when the instantaneous data rate falls below a predefined target rate, denoted by $\hat{R}_k$ for user $k$. This outage probability can be defined as $\text{Pr}[R_k < \hat{R}_k | \Phi_k]$, which represents the probability that the instantaneous data rate $R_k$ falls below the required target $\hat{R}_k$, conditioned on the estimated user location $\Phi_k$. This captures the reliability of the communication link between the AP and user $k$ under imperfect location information.

\subsection{Problem Formulation}
To guarantee the quality of service (QoS) for each user, the outage probability is constrained by a maximum allowable threshold $\varepsilon_o$. That is, the system must satisfy
\begin{equation}
    \text{Pr}[R_k <\hat{R}_k | \Phi_k ] \leq \varepsilon_o, \forall k 
\end{equation}

Under this constraint, the objective is to minimize the total transmit power at the AP, by jointly optimizing the position of the pinching antenna and the transmit power allocated to each user. The optimization problem is formulated as:
\begin{subequations} \label{P_NOMA}
   \begin{align}
    \min_{x^{\text{Pin}}, P_k}~ &\sum_{k=1}^K P_k\\
    \text{s.t.}~& x^{\text{Pin}} \in [0, L], \\
    & \text{Pr}[R_k <\hat{R}_k | \Phi_k ] \leq \varepsilon_o, \forall k, 
\end{align} 
\end{subequations}
where constraint~(\ref{P_NOMA}b) ensures that the pinching antenna remains within the bounds of the dielectric waveguide, and constraint~(\ref{P_NOMA}c) imposes a QoS guarantee for all users by limiting the outage probability.

\begin{figure*}[t!]
    \centering
    \begin{subfigure}[t]{0.5\textwidth}
        \centering
        \includegraphics[width=1\textwidth]{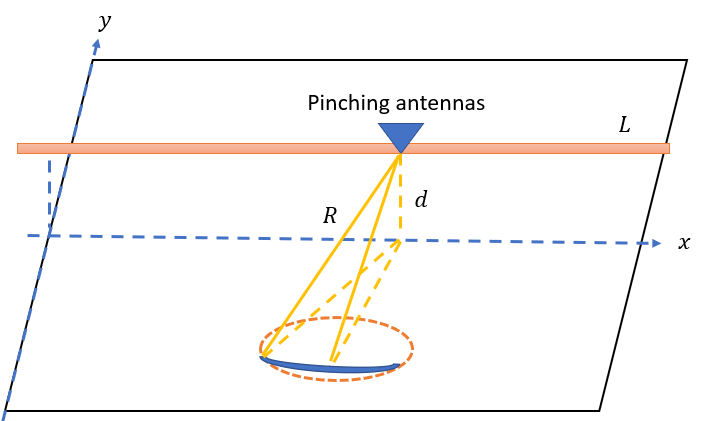}
        \caption{}
        \label{fig:Low_2}
    \end{subfigure}%
    ~ 
    \begin{subfigure}[t]{0.5\textwidth}
        \centering
        \includegraphics[width=1\textwidth]{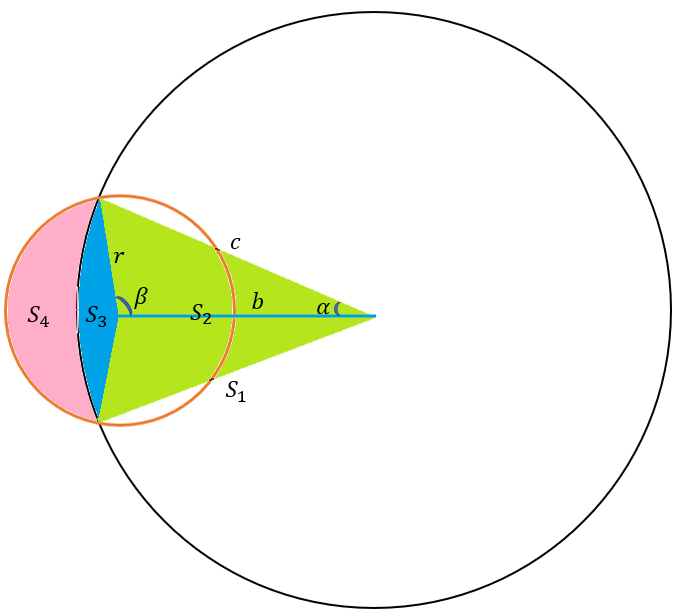}
        \caption{}
        \label{fig:Low_3}
    \end{subfigure}
    \caption {Illustration of a) the intersection between the user circle and pinching antenna sphere; and b) the intersection between the user circle and the circle of the pinching antenna sphere on the ground plane.}
    \label{fig:EE}
\end{figure*}

\section{Proposed Solution}
Problem~\eqref{P_NOMA} is non-convex primarily due to the probabilistic outage constraint in (\ref{P_NOMA}c). Furthermore, the position of the pinching antenna appears in the outage constraint of every user, which introduces interdependence among these constraints. As a result, the outage constraints need to be handled jointly, significantly increasing the complexity of the problem.

To gain insight into how to tackle this challenge, we begin by analyzing a simplified scenario involving a single user. In this case, only one outage constraint is present, which allows us to isolate the key difficulties and develop solution strategies that can later be extended to the general multi-user setting.

\subsection{Single-User Case}
Without loss of generality, we consider an arbitrary user $k$. Under this single-user setting, problem~\eqref{P_NOMA} reduces to the following form:
\begin{subequations} \label{single user}
   \begin{align}
    \min_{x^{\text{Pin}}, P_k}~ &  P_k\\
    \text{s.t.}~& x^{\text{Pin}} \in [0, L], \\
    & \text{Pr}[R_k <\hat{R}_k | \Phi_k ] \leq \varepsilon_o.  
\end{align} 
\end{subequations}

To minimize the transmit power under the given outage constraint, the pinching antenna should be positioned to maximize the user’s channel gain. Given the free-space propagation model, the channel gain is inversely proportional to the squared distance between the user and the antenna. Therefore, the channel gain is maximized when the antenna is located as close as possible to the user. Since the actual user location is uniformly distributed within a circle centered at $(x_k, y_k)$ on the $xy$-plane, the optimal antenna position---minimizing expected distance---should be closest to this center. Specifically, if $x_k > L$, the optimal antenna position is at the boundary, i.e., $x^{\text{Pin}} = L$; otherwise, we set $x^{\text{Pin}} = x_k$, i.e., direct alignment with the estimated $x$-axis coordinate.

We now turn to minimizing the required transmit power without violating the outage constraint. Once the antenna position is fixed, for a given power allocation $P_k$, all user locations that yield the same received signal-to-noise ratio (SNR)---and hence the same data rate, e.g., $\hat{R}_k$---lie on a sphere centered at the antenna with radius $R$. As illustrated by the blue curve in Fig.~2(a), the intersection of this sphere with the user uncertainty circle forms a curve that partitions the uncertainty region into two parts: one closer to the antenna (where $R_k \geq \hat{R}_k$), and the other farther from the antenna (where $R_k < \hat{R}_k$). 
Consequently, the outage region corresponds to the portion of the user circle that lies outside this intersection curve. 
To satisfy the outage constraint, the area of this region must be less than or equal to $\varepsilon_o \pi r^2$, where $r$ is the radius of the user uncertainty circle.

Now the problem lies in how to calculate the area of the outage region. To facilitate this, we will first introduce the following theorem:
\begin{theorem} \label{theorm 1}
    The intersection curve between the antenna sphere and the user uncertainty circle corresponds to an arc with radius $\sqrt{R^2 - d^2}$.
\end{theorem}
\begin{IEEEproof}
    Any point on the intersection curve lies at a distance $R$ from the antenna. Projecting the antenna’s position onto the ground plane yields the point $(x^{\text{Pin}}, 0, 0)$. By the Pythagorean theorem, the distance from this point to any point on the intersection curve is $\sqrt{R^2 - d^2}$, as it forms the base of a right triangle with hypotenuse $R$ and vertical leg $d$. Therefore, the intersection curve is an arc centered at $(x^{\text{Pin}}, 0, 0)$ with radius $\sqrt{R^2 - d^2}$. 
\end{IEEEproof}

According to Theorem~\ref{theorm 1}, we can complete the intersection arc into a circle, and visualize it together with the user uncertainty circle as shown in Fig.~2(b). In this subfigure, the user circle, which represents the uncertainty region of user $k$, is depicted on the left with a radius of $r$. The right circle corresponds to the intersection of the antenna sphere with the ground plane and has a radius of $\sqrt{R^2 - d^2}$. Since the elevation $d$ is typically much smaller than $R$, this radius is generally much larger than $r$, and the right circle appears significantly larger in the figure.
Furthermore, due to the small value of the outage probability threshold $\varepsilon_o$ (e.g., 0.05), the center of the user circle must lie within the antenna circle. This ensures that the pink region, which represents the portion of the user circle where outage occurs, occupies no more than $\varepsilon_o$ of the total area of the user uncertainty region.

Now, the problem lies in calculating the area of the pink region, denoted by $S_4$. Clearly, this area is the difference between the area of the user circle sector and that of the blue region, denoted by $S_3$. The area $S_3$ can be expressed as the difference between the right circle sector area $S_1$ and the area of the two green triangles, i.e., $S_2$.
For notational simplicity, let the radius of the right circle be $c$, such that $c = \sqrt{R^2 - d^2}$. Denote the distance between the centers of the two circles by $b$, where $b = \sqrt{(x^{\text{Pin}} - x_k)^2 + y_k^2}$.

We first compute the area $S_1$. To this end, we need the central angle $2\alpha$ (in radians). According to the law of cosines, the half-angle $\alpha$ is given by
\begin{equation}
   \alpha= \arccos(\frac{b^2+c^2-r^2}{2bc}). 
\end{equation}
Then, the sector area is 
\begin{equation}
    S_1= \alpha c^2 . 
\end{equation}
Next, we calculate the area of $S_2$, which consists of two identical triangles with known side lengths. Let $s$ denote the semi-perimeter of one triangle, i.e., $s=\frac{b+c+r}{2}$. Using Heron’s formula, the area of $S_2$ is given by 
\begin{equation}
    S_2=2 \sqrt{s(s-b)(s-c)(s-r)}. 
\end{equation}
Consequently, we have 
\begin{subequations}
\begin{align}
     S_3&=S_1-S_2 \\
         &=\alpha c^2- 2 \sqrt{s(s-b)(s-c)(s-r)}.
\end{align}
\end{subequations}
Now, we compute the sector area of the user circle, i.e., $S_3 + S_4$. To do this, we calculate the angle $\beta$ (in radians), given by the law of cosines: 
\begin{equation}
   \beta= \arccos(\frac{b^2+r^2-c^2}{2br}). 
\end{equation}
Thus, the total sector area is 
\begin{equation}
    S_3+S_4=(\pi-\beta)r^2,
\end{equation}
and therefore, 
\begin{subequations}
\begin{align}
     S_4&=(\pi-\beta)r^2- S_3 \\
        &=(\pi-\beta)r^2+2 \sqrt{s(s-b)(s-c)(s-r)}-\alpha c^2. 
\end{align}
\end{subequations}
To ensure an outage probability of $\varepsilon_o$, we require
\begin{equation}
    \frac{S_4}{\pi r^2}=\varepsilon_o \rightarrow S_4=\varepsilon_o \pi r^2. 
\end{equation}

Note that in the expression for $S_4$, the only variable is $c$, as $r$ is given and $b$ is determined by the pinching antenna position from the prior optimization step. Due to the complexity of the expression, a closed-form solution for $c$ is intractable. However, it can be verified that $S_4$ is a decreasing function of $c$. Therefore, the bisection method can be used to numerically solve for $c$.
We can set the initial bounds as $b$ and $b + r$. Then, we evaluate the midpoint, $c = b + r/2$. If the corresponding $S_4$ is less than $\varepsilon_o \pi r^2$, the true value of $c$ lies in the right half, and we update the upper bound; otherwise, we update the lower bound. This process is repeated until the interval between the bounds is less than a specified threshold, e.g., $10^{-6}$. Once $c$ is obtained, the corresponding $R$ is calculated as
\begin{equation}
    R=\sqrt{c^2+d^2} . 
\end{equation}

On this basis, the minimum required power can be determined as
\begin{subequations} \label{power_min}
    \begin{align}
        \hat{R}_k&=\log_2 \left(1+ \frac{\eta P_k^{\min}}{  R^2 \sigma_k^2} \right) \\
        \rightarrow P_k^{\min}&= \frac{( 2^{\hat{R}_k}-1 )R^2\sigma_k^2}{\eta}. 
    \end{align}
\end{subequations}

\subsection{Multi-user Case}
Now, let us consider the general multi-user scenario. For a given pinching antenna position, minimizing the total transmit power is equivalent to minimizing the power of each individual user, as user powers are independent. Therefore, the sum power minimization problem can be decomposed into $K$ independent single-user power minimization subproblems. The algorithm proposed earlier for the single-user case can thus be directly applied to determine the minimum power required for each user.

The remaining challenge lies in optimizing the pinching antenna position. The relationship between the minimum required power for each user and the antenna position is governed by the outage probability. However, as shown earlier, this relationship lacks a closed-form expression and is non-differentiable due to the use of the bisection method. As a result, conventional gradient-based optimization methods cannot be applied.

To address this issue, we adopt a heuristic PSO algorithm, which does not require the computation of derivatives. Note that the implementation of the PSO algorithm is standard and readily available in tools such as MATLAB. Therefore, implementation details are omitted. The key component is the evaluation of the objective function, i.e., the minimum sum power, which can be obtained by applying the previously described single-user solution under a given pinching antenna position.

\begin{figure}[ht!] 
\centering
\includegraphics[width=1\linewidth]{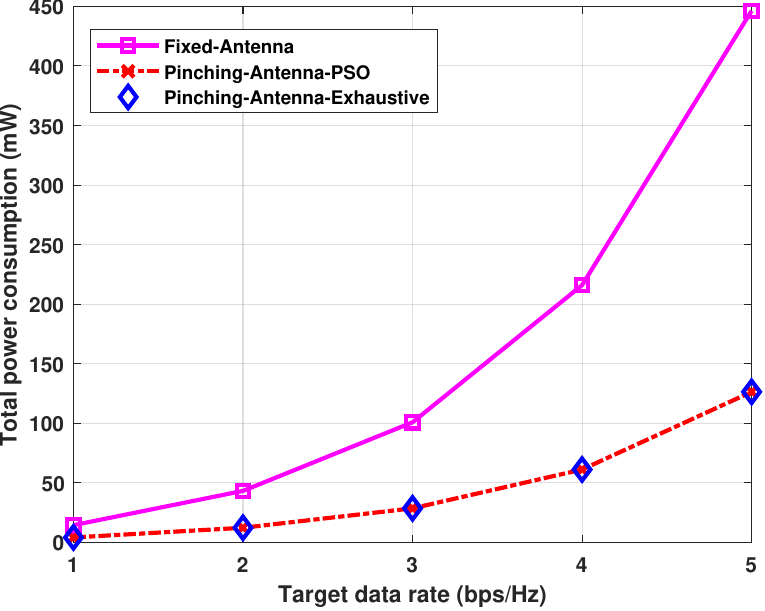}
\caption{Total power consumption versus target rate at the users.} 
\label{fig:Rate}
\end{figure}

\section{Numerical Results}
This section presents numerical results to evaluate the performance of the proposed pinching-antenna-based scheme. The default simulation parameters are as follows: the system serves $K = 5$ users, randomly located within a rectangular service area of length 120 m and width 20 m. The radius of the uncertainty region for each user is set to $r_k = 3$ m, $\forall k$. Each user has a target data rate of $\hat{R}_k = 3$ bps/Hz and a maximum allowable outage probability of $\varepsilon_o = 0.01$, $\forall k$. The carrier frequency is 28 GHz, and the total system bandwidth is 100 MHz. The noise power spectral density is set to $-174$~dBm/Hz. The dielectric waveguide has a length of $L = 50$~m and a height of $d = 3$~m. All results presented are averaged over 1000 independent random realizations.

Two benchmark schemes are considered for comparison:
\begin{itemize}
\item \textbf{Exhaustive search (upper bound):} This approach determines the optimal antenna position via a one-dimensional search along the waveguide.
\item \textbf{Conventional Fixed-antenna architecture:} The antenna is fixed at position $(0, 0, d)$~m, corresponding to a conventional system setup.
\end{itemize}

Fig.~\ref{fig:Rate} illustrates the total transmit power as a function of the target data rate $\hat{R}_k$. As expected, the total power increases exponentially with the target rate across all schemes, consistent with the expression in (\ref{power_min}b), which shows that the minimum required power grows exponentially with $\hat{R}_k$. Among the three schemes, the fixed-antenna-based solution requires substantially more power. In contrast, the proposed scheme closely approximates the performance of the exhaustive search, demonstrating its near-optimality and the significant power savings offered by pinching-antenna systems over traditional architectures.

\begin{figure}[ht!] 
\centering
\includegraphics[width=1\linewidth]{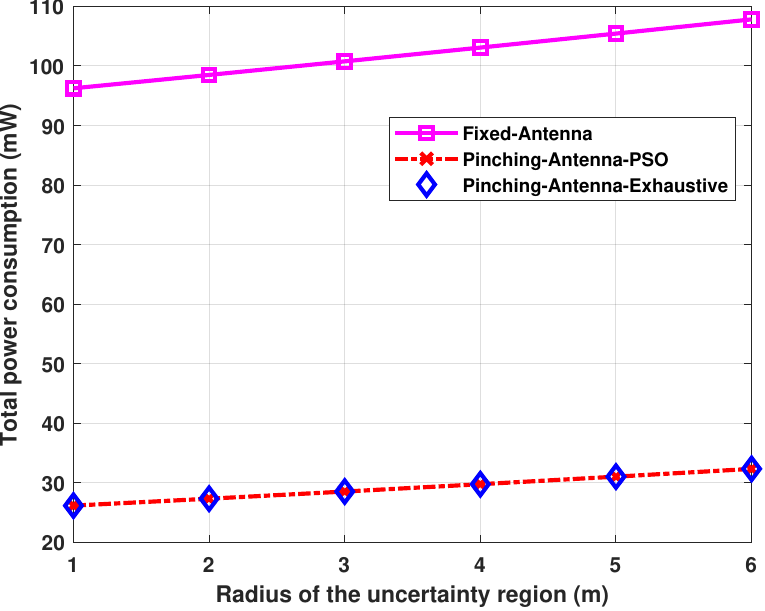}
\caption{Total power consumption versus the radius of the uncertainty region.} 
\label{fig:Radius}
\end{figure}

\begin{figure}[ht!] 
\centering
\includegraphics[width=1\linewidth]{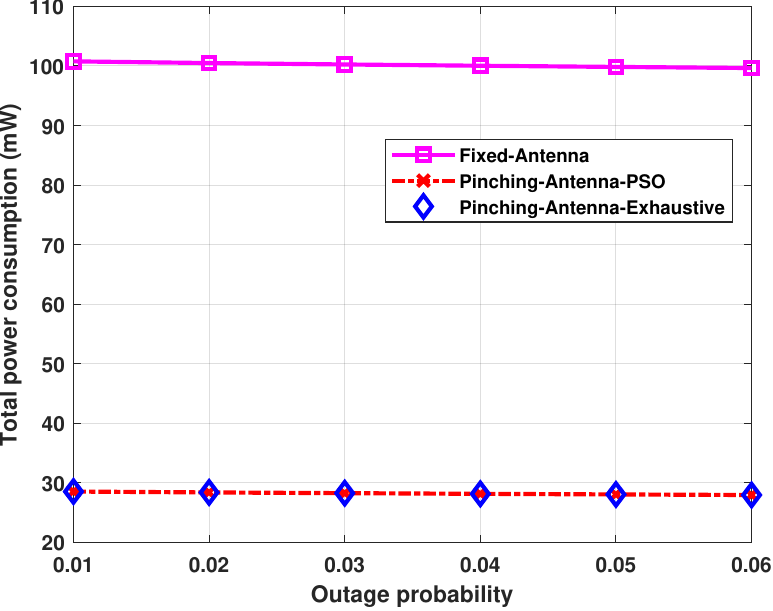}
\caption{Total power consumption versus the outage probability constraint at the users.} 
\label{fig:Outage}
\end{figure}

Fig.~\ref{fig:Radius} shows the impact of the uncertainty region radius $r$ on the total power consumption. As anticipated, increasing the radius leads to higher power consumption in all schemes. The growth is approximately linear, which can be explained as follows: let the value of $c$ obtained via the bisection method be approximated by $c = b + \delta r$, where $\delta \in [0, 1]$ ensures the outage constraint is satisfied. Then, $R^2=c^2+d^2$=$(b+\delta r)^2+d^2 $= $b^2+2b\delta r + \delta^2  r^2+d^2 $. 
Since $b \gg r$ and $\delta \in [0, 1]$, the term $2b\delta r$ dominates over $\delta^2 r^2$, leading to quasi-linear growth of $R^2$ with $r$. As a result, the minimum required power $P_k^{\min} = \frac{(2^{\hat{R}_k} - 1) R^2 \sigma_k^2}{\eta}$ also grows approximately linearly with $r$. The two pinching-antenna-based schemes (proposed and exhaustive) yield nearly identical results and substantially outperform the fixed-antenna baseline.

Finally, Fig.\ref{fig:Outage} depicts the total power consumption as a function of the outage probability threshold $\varepsilon_o$. As expected, total power decreases as $\varepsilon_o$ increases across all schemes. However, the reduction is relatively minor. This is primarily because the uncertainty radius $r$ is significantly smaller than the distance between the circle centers, i.e., $b$. For the considered range $\varepsilon_o \leq 0.5$, the range of $R$ satisfies $b \leq R \leq b + r$, as discussed in SectionIII-B. When $r \ll b$, variations in $\varepsilon_o$ lead to only marginal changes in $R$, and thus, in total power. Once again, the fixed-antenna scheme exhibits considerably higher power consumption compared to both pinching-antenna-based methods.

\section{Conclusion} 
\label{Sec:Conclusion}
This paper studied robust resource allocation for a downlink multi-user system employing a pinching antenna under user location uncertainty. The goal was to minimize total transmit power subject to individual outage probability constraints. Due to the non-convexity of the coupled probabilistic constraints, we first analyzed the single-user case, deriving the optimal antenna position and solving the power minimization via a bisection method combined with geometric analysis. We then extended the resulting power allocation strategy to the general multi-user scenario and optimized the pinching antenna position using the PSO algorithm. Numerical results demonstrated that the proposed scheme achieves near-optimal performance with significant power savings compared to fixed-antenna systems. Future work includes extending the framework to scenarios with multiple pinching antennas or waveguides. 

\bibliographystyle{IEEEtran}
\bibliography{biblio}

\balance

\end{document}